\documentclass[aip,apm,graphicx]{revtex4-1}
\usepackage{amsmath,amsfonts,amssymb,amsthm,graphicx}

\draft 

\begin{document}


\title{A method to extract pure Raman spectrum of epitaxial graphene on SiC\\}



\author{J. Kunc}
\email[]{jan.kunc@physics.gatech.edu}
\affiliation{The Georgia Institute of Technology, Atlanta, Georgia 30332-0430, USA}
\affiliation{Faculty of Mathematics and Physics, Institute of Physics, 12116 Prague, Czech Republic}
\author{Y. Hu}
\affiliation{The Georgia Institute of Technology, Atlanta, Georgia 30332-0430, USA}
\author{J. Palmer}
\affiliation{The Georgia Institute of Technology, Atlanta, Georgia 30332-0430, USA}
\author{C. Berger}
\affiliation{The Georgia Institute of Technology, Atlanta, Georgia 30332-0430, USA}
\affiliation{CNRS/Institute N\'{e}el, BP 166, 38042 Grenoble, France}
\author{W. A. de Heer}
\affiliation{The Georgia Institute of Technology, Atlanta, Georgia 30332-0430, USA}

\date{\today}

\begin{abstract}

A method is proposed to extract pure Raman spectrum of epitaxial graphene on SiC by using a Non-negative Matrix Factorization. It overcomes problems of negative spectral intensity and poorly resolved spectra resulting from a simple subtraction of a SiC background from the experimental data. We also show that the method is similar to deconvolution, for spectra composed of multiple sub- micrometer areas, with the advantage that no prior information on the impulse response functions is needed. We have used this property to characterize the Raman laser beam. The method  capability in efficient data smoothing is also demonstrated.

\end{abstract}


\maketitle 

\section{Introduction}
As graphene is getting more attention as an electronic material~\cite{Berger04}, so is the need for a better evaluation of the sample characteristics. Epitaxial graphene (EG) grown on 4H- or 6H-SiC is a scalable platform for high quality graphene devices on a mono crystalline wafer scale semiconductor. High frequency transistor operation~\cite{Guo13, Moon12}, record spin diffusion lengths for spintronics~\cite{Dlubak12}, large scale transistor integration~\cite{Sprinkle10,Ming} have recently been demonstrated in EG. Smoothed-edge nanoscale ribbons~\cite{Sprinkle10,Ruan12} can also be produced on non-polar SiC facets that present exceptional conducting properties~\cite{Baringhaus12,Ruan12} and wide band-gap semiconducting strips~\cite{Hicks13}.

A widely used characterization of graphene is Raman spectroscopy~\cite{FerrariPRL06, MalardPhysRep473-51-2009}. Unlike graphene transferred to SiO$_2$, the Raman spectra of EG on SiC consists of a combination of signals from graphene and bulk SiC in the spectral region of the graphene D- and G-peaks~\cite{Faugeras,Smet08,Shen08,Ley08,ShivaramanJEM38-725-2009}. This significantly complicates the analysis and data interpretation. SiC and graphene Raman spectra are usually separated by a simple subtraction of a SiC bulk spectrum~\cite{deHeerPhysD10,SuemitsuJournalPhysDAppPhys43-374012-2010}. Subtraction is however poorly defined ~\cite{deHeerPhysD10,SuemitsuJournalPhysDAppPhys43-374012-2010} and may  lead to spectral regions of negative intensity. 
Another issue is related to the Raman spectroscopy of patterned $\mu$m-scale electronic devices. Unless in the near-field regime,  Raman scattering mapping is limited by the spatial resolution given by the laser beam size. Spatial resolution can in principle be increased to the sub-$\mu$m scale by data deconvolution  provided the impulse response function (laser beam shape) is known. 

In this letter, we demonstrate that the Raman spectra of epitaxial graphene can be well decomposed into a pure graphene and SiC spectral parts by using a recently developed Non-negative Matrix Factorization~\cite{Lee99} (NMF) method. In particular, the  resulting graphene Raman spectra are clean, well-resolved and smooth,  even in the spectral range where SiC and graphene peaks overlap. Spatial maps of NMF spectral components correlate with Electrostatic Force Microscopy mapping of differentially graphitized C-face EG samples. We also show that the method can be used for data smoothing. Because NMF does not require a priori information on the impulse response function,  we show that it is a good  alternative to signal deconvolution, and we apply the technique to characterize the spatial spread of the Raman excitation laser beam. 

\section{Samples and experimental techniques}
We have used successfully the NMF in the data analysis of 15 epitaxial graphene samples grown on SiC by the confinement controlled sublimation method~\cite{deHeerPNAS11}. Here we focus on two representative samples, grown on the 4H-SiC(000$\bar{1}$) carbon-face. The first sample is a  C-face multilayer epitaxial graphene (MEG) sample of $\approx$ 5 layers  with homogeneous graphene terraces of $\approx10~\mu$m. For the second sample, MEG was plasma etched away everywhere except in two $10\times10~\mu$m$^2$ MEG areas separated by  a 0.9~$\mu$m wide channel. A sub-single layer graphene was subsequently grown in  the channel and around the MEG areas. This provides three regions on the same sample:  $\approx$ 10 layer thick MEG pads, single graphene layer and bare SiC areas.

Raman scattering was excited by a $\lambda=532$~nm laser light with laser beam size of $\approx1~\mu$m$^2$ (Horiba Jobin-Yvon LabRam). The data were taken between 1000~cm$^{-1}$ and 3600~cm$^{-1}$ with a spectral resolution of 1~cm$^{-1}$. We present below 3 sets of experiments.

\begin{figure}
 \includegraphics[width=7.5cm]{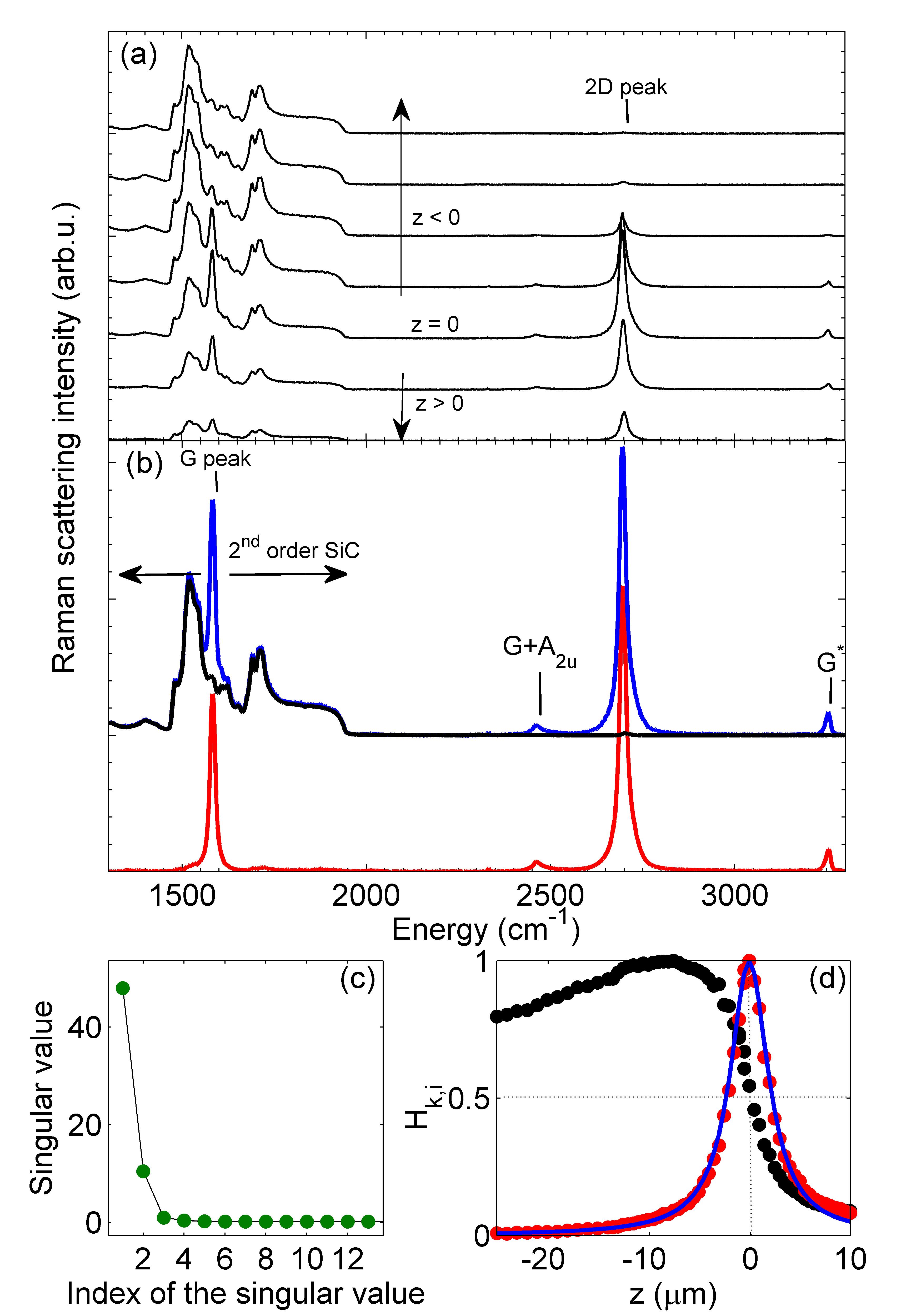}%
 \caption{\label{Fig1} (a) Representative 7 out of 69 Raman spectra used for the NMF decomposition. The focal plane was changed from $z=10~\mu$m above ($z>0$)  down to $z=-25~\mu$m below ($z<0$) the sample surface. (b) Black (SiC) and red (graphene) decomposed spectra are compared with blue Raman spectrum at $z=0~\mu$m before decomposition. (c) First largest singular values of 69 Raman spectra used for NMF decomposition. (d) Linear coefficients (relative strengths) of graphene (red)  and SiC (black) basis functions versus focal plane distance from sample surface. The blue curve is a fit assuming Gaussian beam of the excitation laser (Rayleigh parameter $z_R=2.3~\mu$m).} %
 \end{figure}

In the first experiment (Fig.~\ref{Fig1})  we demonstrate the  decomposition of the  Raman spectra from the first MEG sample onto a SiC and a graphene parts. Each of the $m=69$ Raman spectra was taken at a different focal plane of the laser beam between $z=10~\mu$m above the surface down to $z=-25~\mu$m below, in steps of $\Delta z=0.5$ to 1~$\mu$m.  The focal plane can be adjusted with respect to the sample surface with a precision of $\pm0.25~\mu$m. This allows to change the relative strength of the graphene and SiC signals. The exact $z$ position is not important as long as a sufficient number of spectra are taken in and deep below the graphene plane (where there graphene or SiC signal prevails, respectively). 

In the experiment on the second sample, we identify the 3 regions of various graphene thickness. In this case we acquired $m=828$ spectra by mapping the surface at constant $z=0~\mu$m. The surface area $3.4\times9~\mu$m$^2$ was scanned with a latteral resolution 0.2~$\mu$m (18 and 46 steps in $x$ and $y$ latteral direction, respectivelly).

The NMF is a multivariate analysis tool providing data categorization as K-means~\cite{SchmidtPRB84-235422-2011} and spectral decomposition as Principal Component Analysis (PCA). In contrast to PCA, which decomposes data into orthonormal spectral basis, NMF imposes non-negativity constraint, thus facilitating the interpretation of basis spectral functions. We use freely accessible implementation of NMF~\cite{JinguKim08,HyunsooKim08,ParkNMFcode} using both Block Principal Pivoting and Active Set~\cite{HyunsooKim08} minimization algorithms. Beside non-negativity constraints, constraints on sparsity~\cite{Hoyer04} and regularity~\cite{JinguKim08} of basis functions and/or linear coefficients have been applied.

In both experiments all the spectra were normalized by their global maximum (maximal Raman scattering intensity within all 69 or 828 spectra, respectivelly) and a local constant background was subtracted from each spectrum separatelly. The data were organized as columns of matrix $V_{exp}$. The matrices $V_{exp}$ were decomposed by NMF giving elementwise non-negative matrices $H$ and $W$ such that $V_{exp}=WH+E$. The error matrix $E$, which Frobenius norm is minimized by the NMF algorithm, contains the noise information and the matrix $V=WH$ the smoothed experimental data. Data smoothing by NMF is discussed at the end of this letter (Fig.~\ref{Fig3}). 
The columns of matrix $W$ are the basis functions (the EG and SiC spectra for instance) and rows of matrix $H$ are the corresponding linear coefficients describing the basis function weight in each experimental spectrum. The matrices W and H are therefore of rank $n\times k$ and $k\times m$, respectively, where $k$ is the number of basis functions, $n$ the number of data points in each experimental spectrum and $m$ the number of experimental spectra. 
The number of basis functions  $k$ is determined by the number of largest singular values of the matrix $V_{exp}$ factorized by Singular Value Decomposition (SVD). 
\section{Results}
\subsection {SiC-EG spectral decomposition - MEG sample}
A few representative Raman spectra as a function of the relative position $z$  for the first MEG sample are shown in Fig.~\ref{Fig1}~(a). Each spectrum has a different SiC vs graphene weight. When the excitation beam is focused at the sample surface  ($z=0~\mu$m in Fig.~\ref{Fig1}~(a) and blue curve in Fig.~\ref{Fig1}~(b)), the Raman spectrum consists of a combination of 2$^{nd}$ order bulk SiC~\cite{WindlPRB49-8764-1994,BurtonPRB59-7282-1999} and  G, G+A$_{2u}$, 2D and G$^{*}$ graphene Raman peaks~\cite{MalardPhysRep473-51-2009,KraussPRB79-165428-2009}. The missing graphene D-peak indicates  good quality graphene. Note the single 2D graphene peak indicating that the 5 layers in the  MEG sample are not stacked like in graphite, as previously discussed~\cite{Faugeras, deHeerPhysD10}. The overall Raman signal intensity is attenuated when the focal plane is high above the sample ($z>0$) and increases when approaching  it ($z\rightarrow 0~\mu$m). Lowering the focal plane below surface leads to attenuation(enhancement) of the graphene(SiC, resp.) signal. The remaining SiC signal gets slightly reduced for $z<-10~\mu$m (below the surface) due to absorption of excitation laser beam in SiC substrate. 

The first 13 principal values of matrix $V_{exp}$ obtained by SVD  are plotted in Fig.~\ref{Fig1}~(c). 
There are two principal components represented by the two largest principal values, as seen in Fig.~\ref{Fig1}~(c), the other singular values have negligible contribution. The matrix $V_{exp}$ is then factorized with NMF of rank $k=2$, giving two non-negative basis functions (columns of matrix $W$) plotted in red and black in Fig.~\ref{Fig1}~(b). The basis functions can be attributed to  the SiC and graphene Raman spectra. The linear coefficients (the 2 rows of matrix $H$, plotted in Fig.~\ref{Fig1}~(d)), show the relative contributions of the SiC (black dots) and graphene (red dots) to the measured Raman spectra (Fig.~\ref{Fig1}~(a)) as a function of $z$. As expected the graphene signal is maximized for $z=0$ i.e. when the excitation laser beam is focused at the sample surface. 

Assuming a Gaussian shape for the laser beam\cite{TeichSaleh}, the $z$-dependence of the graphene signal can be fitted by $H_{1,i}=\frac{1}{1+(z/z_R)^2}$. Here $H_{1,i}$ is the row of linear coefficients for graphene as determined by NMF, $z_R=\frac{\pi w_0^2}{\lambda}=2.3~\mu$m is a Rayleigh parameter of the Gaussian beam and $2w_0=1.2~\mu$m is the beam waist (i.e. the radial width). The beam Full Width at Half Maximum FWHM=1.2~$\mu$m is in agreement with expected $\approx 1~\mu$m (microscope objective numerical aperture $\mathrm{NA}=0.90$). 

Note that for performing spectral decomposition only, the number of spectra (here $m=69$) can be greatly reduced, as long as the condition $(n+m)k<nm$ is fulfilled~\cite{Lee99}  (here $n=1867$ data points per spectrum ). The large $m$ was motivated here to have high enough spatial resolution for the laser beam characterization.
 \begin{figure}
 \includegraphics[width=7.5cm]{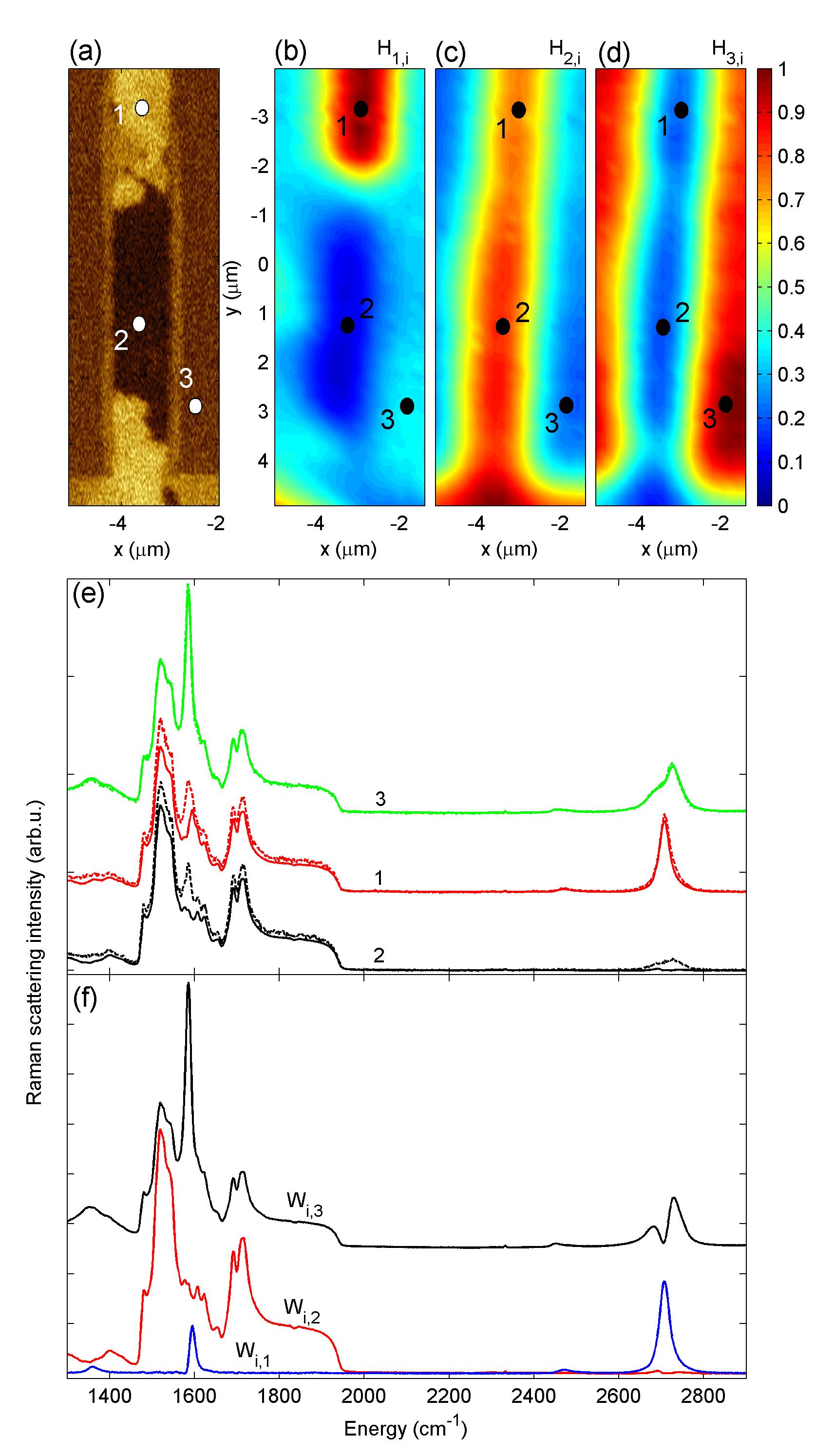}%
 \caption{\label{Fig2} (a) Electrostatic force microscopy image. Maps of the linear coeficients of (b) single layer grahene, (c) SiC and (d) multilayer graphene of the same area on the sample. (e) Normalized data (dashed lines) measured at points 1, 2 and 3 (shown in (a)-(d) by white and black numbered circles) are compared to NMF filtered spectra (solid lines) using threshold 0.3. (f) Blue, red and black spectra are basis functions corresponding to linear coefficients in (b), (c) and (d), respectively.}%
 \end{figure} 
 \subsection {Spectrum deconvolution  - MEG patterned sample}
We now show how NMF can be used as an alternative to deconvolution. The second (patterned) sample  was mapped with Atomic Force Microscopy (AFM), Electrostatic Force Microscopy (EFM) (Fig.~\ref{Fig2}~(a)) and Raman scattering. Three different graphene/SiC areas are identified  in the EFM image (Fig.~\ref{Fig2}~(a)).  The brown areas in Fig.~\ref{Fig2}~(a) (area 3) are the MEG pads, the light areas (such as 1) are single layer graphene, the dark one (labelled 2) is non-graphitized SiC substrate. The normalized Raman spectra measured at points 1, 2, 3 are plotted as dashed lines in Fig.~\ref{Fig2}~(e). The D, G and 2D graphene peaks are clearly identified in the three spectra, as well as the SiC Raman peaks. 
The 2D peak in area 3 presents a double peak structure which is some times observed in MEG sample. It can be due to the contribution of AB stacked layers (about 10-15\% of the stacking \cite{HicksJPhysDApplPhys45-2012}) or small  graphitic areas \cite{Faugeras}, that might be more common in this re-annealed sample.

Because we clearly identified  three regions,  the  Raman spectra  at each point of the mapping were decomposed by NMF of rank $k=3$. The three basis functions $W_{i,1}, W_{i,2}, W_{i,3}$ (columns of matrix $W$) are plotted in Fig.~\ref{Fig2}~(f). The $W_{i,1}$ and $W_{i,2}$  basis function can clearly be associated with single layer graphene and bare SiC, respectively, whereas  $W_{i,3}$  is more complex.  
The linear coefficients $H_{1,i}, H_{2,i}, H_{3,i}$ (rows of matrix $H$) plotted in Fig.~\ref{Fig2}~(b-d) show that each basis function $W_{i,1}$, $W_{i,2}$ and $W_{i,3}$ clearly dominates in a different area of the sample. Particularly area 2 shows well pronounced minimum for both graphene related basis functions $W_{i,1}$ and $W_{i,3}$. 

Because the channel (0.9~$\mu$m wide) is narrower than the laser beam spot  (1.2~$\mu$m, as determined above) the Raman signal coming from the channel (areas 1 and 2) gets mixed with the signals from the MEG pads. We now show that the Raman spectra of the narrow channel can nevertheless be reconstructed, by keeping only the large contributions to the Raman signal in each area.
For this we performed an inverse data composition $V=WH$ by  replacing the linear coefficients in matrix $H$ by $H'$, where $H'_{ij}=H_{ij}$ for $H_{ij}>H_{th}=0.3$ otherwise $H'_{ij}=0$. Applying this threshold on the inverse data composition gives the filtered Raman spectra of  Fig.~\ref{Fig2}~(e) (solid lines). The graphene related signal in area 1 is still present, however the D, G and 2D peaks are filtered out at point 2. Changing the threshold $H_{th}$ sets the sensitivity of the filtering. At low(high) value of $H_{th}$ the measured data (basis functions), resp. are retrieved. Hence, the threshold $H_{th}$ effectively removes non-local spurious signal  and plays a similar role as deconvolution. Importantly though, no information on a impulse response function is needed.
The filtered Raman spectra prove that area 1 is a single layer as shown by its weak G-peak and narrow  Lorentzian 2D peak (FWHM=27~cm$^{-1}$). Area 3 is covered by few layer graphene (strong G-peak, splitted 2D peak, that may come from layers of different strain or doping in the MEG stack ~\cite{MalardPRB76-201401-2007,MalardPhysRep473-51-2009}  and attenuated intensity of SiC Raman scattering~\cite{ShivaramanJEM38-725-2009}), as expected from the growth conditions. Area 2 is bare SiC, showing that the regraphitization process in the trench is only partial. 

We note that the NMF spectral decomposition is not unique. Each decomposition $V=WH$ can be replaced by $V=\tilde{W}\tilde{H}=WD^{-1}DH$, for any non-negative regular square matrix $D$. The non-uniqueness can be treated, among others, by additional constraints on the matrices $W$ or $H$. The most common constraints are sparsness~\cite{Hoyer04,JinguKim08} and regularity~\cite{JinguKim08}.  This ambiguity of NMF is also reflected in our analysis. For instance, the basis function $W_{i,3}$  (Fig.~\ref{Fig2}~(f)) is a mixture of graphene and SiC Raman signals. The mixed basis function $W_{i,3}$ can be further decomposed either by applying different NMF constraints, by finding a transformation matrix $D$ or by a Raman depth analysis, as discussed above (Fig.~\ref{Fig1}). 

 \begin{figure}
 \includegraphics[width=7cm]{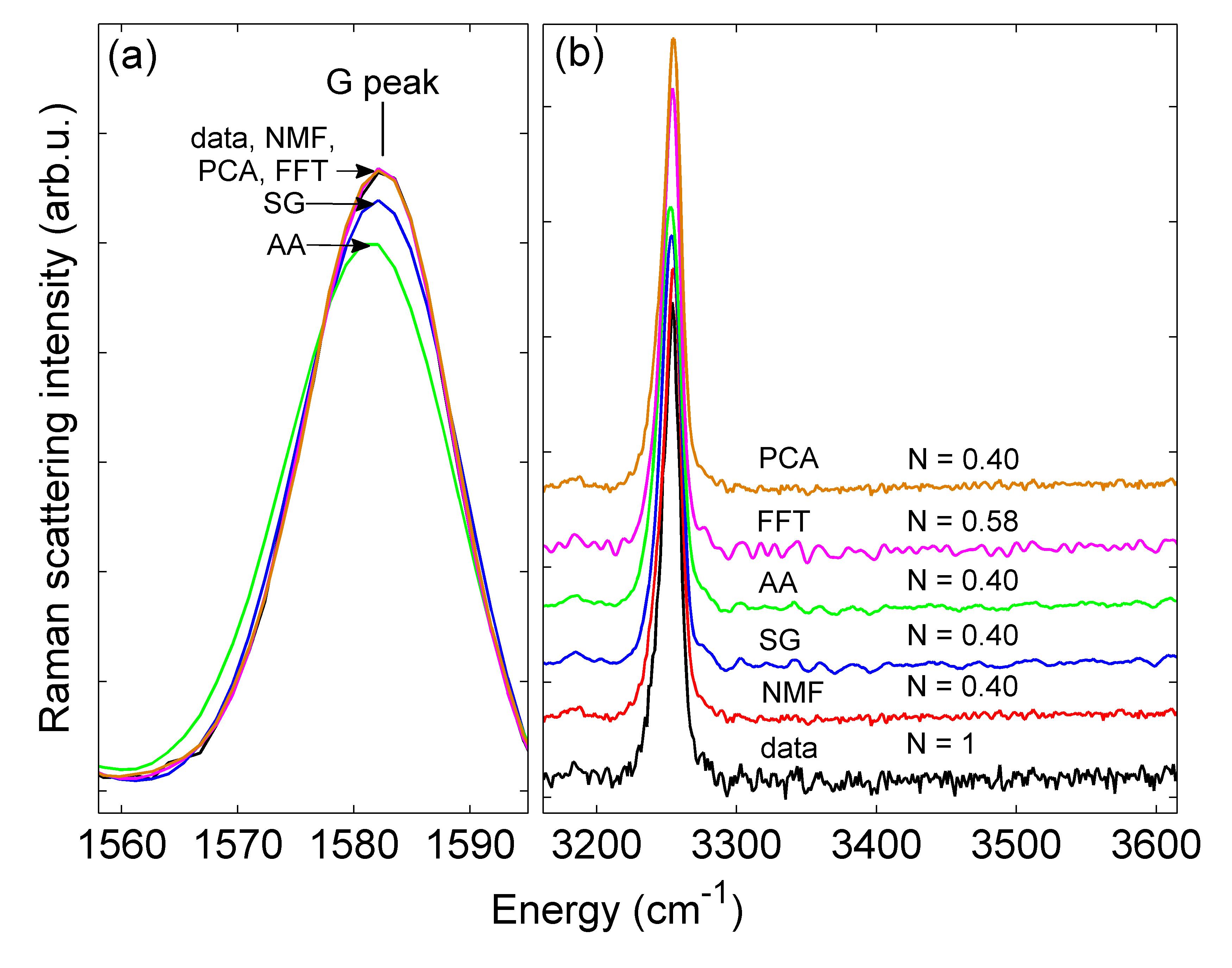}%
 \caption{\label{Fig3} Raman spectra of epitaxially grown graphene on SiC. NMF is used for data (black curve, marked as data) smoothing instead of decomposition. Several smoothing algorithms are compared. Red, blue, green, magenta and yellow curves show smoothed data using 2 component NMF, $2^{nd}$ order Savitzky-Golay with widnow lentgth 15 points (SG), adjacent averaging using 8 point rectangular smoothing window (AA), high-pass Fouier transform filter (FFT, high-pass frequency cut off 0.12~cm) and 2 component Principal Component Analysis (PCA) algorithms. Influence of data smoothing on the (a) G-peak intensity and (b) flat part of the spectrum is shown.}%
 \end{figure}

\subsection {Data smoothing - MEG sample}
Next, we use NMF for data smoothing. The exact NMF data decomposition is $V_{exp}=WH+E$, where the Frobenius norm of matrix $E$ is minimized by the NMF algorithm. Therefore matrix $E$ contains mainly the noise and the product $V=WH$ carries the smoothed experimental data. We use the Raman spectra measured on the first MEG sample to compare NMF smoothing with four most frequently used smoothing algorithms  (see Fig.~\ref{Fig3}). We define the normalized data noise level by the ratio $N=N_{sm}/N_{exp}$ of the standard deviation of the smoothed data $N_{sm}$ to the standard deviation of the experimental data $N_{exp}$. The smoothing parameters of Savitzky-Golay (SG), adjacent averaging (AA) and Fast Fourier Tranform high pass filter (FFT) are set so that the data noise level is reduced to the level of NMF 2-component smoothing $N=0.4$. We compare the noise level in a region of the spectrum with no peak  in Fig.~\ref{Fig3}~(b) and the  intensity of the sharp G-peak (that sits on top of the SiC peaks) in Fig.~\ref{Fig3}~(a). SG and AA smoothing already influence the intensity of G-peak and high pass FFT filter does not reduce the noise level below $N=0.58$. Worse, slowly varying parts of the Raman spectrum are modified also when the low-pass FFT filter is used. PCA and NMF smoothing both reduce the noise level to $N=0.4$ and keep the sharp G-peak intensity. Hence, both NMF and PCA provide the most efficient noise reduction without modifying any information contained in a Raman spectrum of graphene epitaxially grown on SiC.

\section{Conclusions}
The Non-negative Matrix Factorization was successfully applied to decompose the Raman scattering spectra of epitaxial graphene into a graphene and a SiC components. In contrast to SiC background subtraction, the method provides well defined and non-negative decomposed spectra. Corresponding linear coefficients carry additional information on the spatial origin of the signal. This was used to determine the excitation laser beam size and to filter  Raman spectra from spurious signal in a mapping of sub-beam size EG features. We have shown that NMF can be used as an alternative to deconvolution and NMF data smoothing capabilities were demonstrated.

\begin{acknowledgments}
 We acknowledge financial support from  NSF-MRSEC (\#DMR-0820382),  AFSOR, the W.M. Keck foundation and the Partner University Fund.
  \end{acknowledgments}

%


\end{document}